 \documentclass[final,5p,times,twocolumn,authoryear]{elsarticle}

\usepackage{amsfonts,amsmath,amssymb,amsthm}

\usepackage{graphicx}
\usepackage{dcolumn}
\usepackage{bm,dsfont}

\usepackage{titlesec}

\usepackage{hyperref}

\hypersetup{
	pdfnewwindow=true,      
	colorlinks=true,       
	linkcolor=blue,          
	citecolor=blue,        
	filecolor=blue,      
	urlcolor=blue,          
}

\usepackage{color}

\usepackage{lineno,gensymb,lipsum}
\makeatother
\usepackage{ulem}

\newcommand{\be}{\begin{equation}}
\newcommand{\ee}{\end{equation}}

\usepackage{braket}
\usepackage{mathtools,siunitx}
\usepackage{changes}
\definechangesauthor[name={qqgu}, color=orange]{qg}
\usepackage{ulem}

\begin{document}

\begin{frontmatter}


\title{Pitfalls of Exchange-Correlation Functionals in Descriptions of Magnetism:\\
  Cautionary Tale of the FeRh Alloy}
\author[first]{Shishir Kumar Pandey}\ead{shishir.kr.pandey@gmail.com}
\author[second]{Saikat Debnath}
\author[first,third]{Zhanghao Zhouyin}
\author[first,fourth]{Qiangqiang Gu}\ead{guqq@pku.edu.cn}

\affiliation[first]{organization={AI for Science Institute},
            city={ Beijing},
            country={China}}
\affiliation[second]{organization={M. V. College},
            city={Buxar},
            state={Bihar},
            country={India}}
\affiliation[third]{organization={College of Intelligence and Computing, Tianjin University},
            city={Tianjin},
            postcode={300350}, 
            state={Bihar},
            country={China}}
\affiliation[fourth]{organization={School of Mathematical Science, Peking University},
            city={ Beijing},
            postcode={100871}, 
            country={China}}
            
%



%

\date{\today}

\begin{abstract}
The magnetic ground state of FeRh is highly sensitive towards the lattice constant. This, in addition to partially filled d-shells of Fe and Rh, posed a significant challenge for Density Functional Theory (DFT) calculations in the past. Here, we have investigated the performance of various exchange-correlation (XC) functionals within the DFT formalism for this challenging binary alloy. We have employed Local Spin Density Approximation (LSDA), various Generalized Gradient Approximations (GGAs), and newly developed Strongly Constrained and Appropriately Normed (SCAN) meta-GGA functional. Our results show the limitations of any single functional in capturing the intricate interplay of structural, electronic, and magnetic properties in FeRh. While SCAN can accurately describe some magnetic features and phonon dispersion, it significantly overestimates the Fe-Fe magnetic interactions, leading to an unreasonable magnetic ordering temperature. Conversely, the Perdew-Burke-Ernzerhof (PBE) GGA exhibits the opposite behavior. These findings highlight the challenges in simulating materials with partially filled $d$-shells using DFT, underscoring the crucial need for developing a versatile XC functional that can effectively account for the multifaceted nature of such systems.

\end{abstract}


\end{frontmatter}

\section{Introduction}

Transition metal compounds have become a focal point of intense research due to their remarkable and exotic properties.  Some examples include various forms of magnetism~\cite{my1, my2, my3, my4, my5, my6, my7, my8} and superconductivity~\cite{my9, rmpmott} and different kinds of phase transitions driven by external perturbations such as doping~\cite{my10, rmpmott} and temperature. Magnetic transitions in these compounds are typically second-order, characterized by a continuous evolution from one magnetic state to another without any discontinuous changes in structural parameters. While first-order transitions can occur, their presence remains relatively rare. In these magnetostructural first-order phase transitions, a simultaneous change in the magnetic and structural phase of the material with external stimuli like temperature, magnetic fields, or strain originates from a strong coupling between lattice and magnetic degrees of freedom. The purported technological potential of these first-order magnetic transitions in magnetic recording~\cite{mag_reco} magnetocalorics~\cite{mag_cal}, spintronics~\cite{spintronics}, magnetorestrictions~\cite{magnetorestriction}, heat-assisted magnetic recording~\cite{hamr}, antiferromagnetic (AFM) spintronics~\cite{afm_spintronics1, afm_spintronics2}, and room-temperature AFM memory resistor~\cite{mem_regi} makes them even more fascinating.

One particularly interesting example of such magnetostructural phase transition occurs for a transition metal binary alloy, FeRh~\cite{ferh1, ferh2}.  BCC-based (CsCl-type) ordered phase of FeRh, upon lowering the temperature, first goes through a paramagnetic to ferromagnetic (FM) second-order phase transition at $\sim$ $T_\text{C}$ = 670 K. Further 
lowering of temperature then leads to an FM to AFM transition at $\sim$ $T_\text{N}$  = 370 K. A sudden increase in resistivity and $\sim$ 1\% volume collapse with lowering of temperature in AFM phase reveals first-order nature of the AFM-FM transition~\cite{ferh2}. The reversible nature of this phase transition has also been investigated with the application of an external magnetic field both in bulk and thin films~\cite{ferh_mag_fld2, ferh_mag_fld}. What makes this material particularly interesting in the context of device applications is its AFM-FM transition being close to the room temperature.

Efforts to investigate phase transitions in FeRh have yielded conflicting perspectives, with some researchers suggesting simultaneous structural and magnetic changes~\cite{dispute1, dispute2} while others contend that the magnetic transition precedes unit cell expansion~\cite{dispute3, dispute4, dispute5, dispute6, dispute7}. To gain a deeper understanding of such scenarios, Density Functional Theory (DFT)-based calculations are often employed for their precision and reliability. However, in the case of FeRh, prior DFT-based calculations indicate an AFM-FM transition at a significantly larger volume~\cite{dft_lat1, dft_lat2, dft_lat3, dft_lat4, dispute5, dft_lat5} compared to the experimentally observed 1 \% change.  Despite the magnetic moments of Fe/Rh ions, obtained in these calculations, aligning well with experimental data~\cite{dispute3, dispute6, dispute8, dispute9, dft_lat1, dft_lat3, dft_lat5,  dft_lat6}, it is surprising that DFT calculations fail to capture the AFM-FM transition. In instances where DFT-derived parameters are used for modeling, there is a risk of reaching inaccurate conclusions. Therefore, a reassessment to identify the issues with DFT calculations in this system is warranted.

Central to these DFT calculations are exchange-correlation (XC) functionals such as local spin density approximation (LSDA) or various generalized gradient approximations (GGAs). It has become nearly established that these functionals inadequately treat the correlated $d$ orbitals of transition metals, leading to issues in describing materials like strongly correlated superconductors~\cite{failed_d} and nickel oxide~\cite{failed_d2}.
Such erroneous description arises mainly due to self-interaction error~\cite{self_intn} often causing inconsistency between theory and experiments. In this context, alternative approaches like DFT+$U$~\cite{hubbu, ldau} and semi-local meta-GGA density functionals such as strongly constrained and appropriately normed (SCAN) functional~\cite{scan1, scan2} have demonstrated notable success in describing the properties of various materials~\cite{scan_app1, scan_app2, scan_app3, scan_app4}. SCAN, in particular, excels in capturing the XC energy of atoms and is computationally efficient compared to accurate hybrid functionals~\cite{hybrid}. Therefore, it is worthwhile to investigate whether the previous discrepancies of DFT calculations for FeRh can be mitigated by employing different XC functionals. If so, understanding the extent of the differences in results across various functionals is crucial. It's noteworthy that this issue has also been recently  highlighted in a comprehensive DFT study by Bosoni {\it et al.}~\cite{nat_2023_dft}. 

In this study, we conducted a comparative analysis of results obtained from the meta-GGA SCAN functional, LDA, and various GGA functionals. We find that SCAN provides a better description of volume expansion accompanying AFM-FM transition and magnitude of magnetic moments of Fe/Rh atoms in FeRh than any other functional. While the phonon dispersions from SCAN and revised-PBE (RPBE) exhibit close agreement, there are qualitative differences in the spin-polarized band structures between the two cases. We attribute this disparity to the larger exchange splitting observed in the band structure calculations using SCAN.
However, magnetic interactions between Fe atoms ($J^\text{Fe-Fe}$) in the AFM ground state, calculated using the Frozen Magnon approach, are highly overestimated resulting in unreasonably large magnetic ordering temperature in our Monte-Carlo simulations. In this scenario, the PBE functional performs reasonably well. Our results emphasize that the intricately intertwined structural, electronic, and magnetic properties of FeRh cannot be accurately described by a single functional. This underscores the need for further development of a versatile exchange-correlation functional that can effectively capture the multifaceted nature of such materials.

\section{Methodology}

\subsection{Total energy and phonon calculations}

Density-functional theory calculations are performed using projector-augmented wave method \cite{PAW,PAWpotentials1}, implemented within Vienna $ab$ $initio$ simulation package (VASP)~\cite{Kresse}. Total energy calculations are done with 
LDA,  GGA-based functionals namely PBE~\cite{PBE}, revised PBE (RPBE)~\cite{rpbe}, revised PBE for solids (PBEsol)~\cite{pbesol}, and meta-GGA functional namely SCAN~\cite{scan1}. In GGA pseudopotentials of Fe and Rh, 3$d^7$4$s^1$ and 4$p^6$4$d^8$5$s^1$ are treated as valence states. 
 We start with a sixteen-atom conventional magnetic unit cell of cubic crystal structure (space group: $P\bar{m}3m$, no: 221) of FeRh made from the two-atom primitive cell with lattice constant 2.99 \AA{}. Given the importance of spin-lattice coupling in this material, full optimization of crystal structure was performed with each of the pseudopotentials after imposing the G-type AFM ground state (every Fe-Fe first neighbors are antiferromagnetically coupled) observed experimentally at low temperatures. Energy and Hellmann-Feynman force convergence criteria of 10$^{-7}$ eV and 10$^{-4}$ eV/\AA{} respectively are considered during the optimization with the aim of phonon calculations in the next step. Plane-wave cutoff energy 550 eV and 9 $\times$ 9 $\times$ 9 $\Gamma$-centered $k$-mesh, Gaussian smearing width of 0.05 eV are considered in our calculations. The results obtained are well converged with respect to these parameters. 
 
 Structural optimizations with different XC-functional lead to quite small changes in the lattice constants except for the LDA as shown in Table~\ref{tab:lat}. The total energy of the FM state and phonon dispersion for both AFM and FM states are calculated with PBE, RPBE, and SCAN functionals with respective optimized structures. We employed the finite displacement method in phonon calculations and the post-processing is done using Phonopy~\cite{phonopy,phonopy2} software. 

\subsection{Estimation of magnetic interactions and Monte Carlo simulations}

To quantify the role of XC functionals on the magnetic ground state, exchange interactions between Fe atoms in the AFM ground state are estimated with the Frozen-magnon approach (FMA) for both PBE and SCAN functionals.  In this method, the spin spiral state is represented with propagation vector (\textit{\textbf{q}}) defined in the reciprocal space of the lattice. Each spin spiral state is equivalent to a magnetic configuration characterized by, 
$$
\bm{e_i} = \left( \sin{\theta}\cos{(\bm{q}\cdot \bm{R}_i)}, \sin{\theta}\sin{(\bm{q}\cdot \bm{R}_i)}, \cos{\theta} \right)
$$
where $\bm
{R}_i$ = $\bm{R} + \bm{\tau}_\nu$ = $\bm{R}_{\nu+\bm{R}}$, and $\bm{R}$, $\bm{\tau}_\nu$ are the lattice and the basis vectors respectively. $\theta$ is the polar angle made by the spin with the $z$ axis. Using noncollinear magnetism formulation implemented within the VASP package, energy corresponding to each spin-spiral state $E(\bm{q})$ is calculated with 16 atom AFM unit cell. Assuming a localized spin at the Fe site, the results are then mapped onto an extended classical Heisenberg Hamiltonian,

\begin{equation}
   H = -\sum_{i \ne j} J_{ij} \bm{e_i} \cdot \bm{e_j}
\end{equation}

where $J_{ij}$ is the exchange interaction between the Fe spins at sites $i$ and $j$, $\bm{e_i}$ and $\bm{e_j}$ are the unit vectors at these two sites. Within this Heisenberg model, the energy of any frozen-magnon configurations represented by $\bm{q}$ can be given by,
$$
\ E(\theta, \bm{q}) = E_0(\theta) - \sin^2(\theta) J(\bm{q}) 
$$
For simplicity, in our calculations, we restricted the spins to lie in the $xy$ plane.
Fourier transform of $J_{ij}$ in the above equation can be defined as, 
$$
J^{\mu \nu}(\bm{q})  = \delta_{\mu \nu} J_{\mu \mu} - \sum_{\bm{R}} J_{\mu [\nu+\bm{R}]} e^{-i({\bm{\tau}_\mu-\bm{\tau}_\nu-\bm{R}})\cdot \bm{q}}
$$
By taking the inverse Fourier transform, we can obtain, the real space interaction parameters as,  
$$
J_{ij} = \frac{1}{N}\sum_{\bm{q}} J^{\mu_i \nu_j} (\bm{q})e^{i({\bm{\mu}_i-\bm{\nu}_j-\bm{R}})\cdot \bm{q}}
$$
The magnitude of spins is absorbed in the obtained $J$'s which are then used in classical Monte Carlo (MC) simulations to estimate the magnetic ordering temperature. In our in-house MC code, unit vectors on a simple cubic lattice of 16 $\times$ 16 $\times$ 16 are considered, and starting from a
random spin configuration, the system is brought into a thermal equilibrium within 2 $\times$ 10$^5$ MC steps at each temperature. After the thermal equilibrium is achieved, we calculated the magnetization of each sublattice which was further used to determine the magnetic ordering temperature. Our results are well-converged with respect to both the lattice size and the number of MC steps. 

\section{Results}

Our analysis begins with the optimization of the crystal structure. For materials exhibiting magnetostructural transitions, a strong lattice-spin coupling is expected. In the context of materials undergoing magnetostructural transitions, a robust lattice-spin coupling is anticipated. In such scenarios, a comprehensive structural optimization proves invaluable for gaining insights. Utilizing various exchange-correlation (XC) functionals, we systematically optimize the crystal structure while considering the antiferromagnetic (AFM) state, as detailed in the methodology section. 

Table~\ref{tab:lat} presents the optimized lattice constants, their changes observed during optimization, and the magnetic moments of Fe atoms. Throughout these calculations, the crystal structure maintains its cubic symmetry. Consistent with findings from prior Density Functional Theory (DFT)-based studies, the magnetic moments of Rh atoms persist at zero in the AFM ground state. 

\begin{table}[!ht]
 \centering
 \caption{A comparison of the lattice constants, their variations, and the magnetic moments of Fe after optimizing the crystal structure using different XC functionals for the AFM state. Additionally, the difference in total energy (per formula unit) with the ferromagnetic (FM) state is provided. Notably, the results obtained with the SCAN functional stand out distinctly from those obtained with other XC-functionals.}
\begin{tabular}{cccccc}
\hline 
 &  &   &    &   &    \\
XC-functional & & $a$ (\AA{})  & $\delta a$ (\%)  & $\mu_B^\text{Fe}$ & E$_\text{FM}$-E$_\text{AFM}$ \\
 &  &   &    &   &   (meV)  \\
\hline
  &  &   &    &   &   \\
LDA    &&  5.832 &  ~-2.3  &  2.810   & 92 \\ 
PBE    &&  5.979 &  +0.2  &  3.078    & 66 \\
RPBE   &&  6.031 &  +1.0    &  3.145  & 46 \\  
PBEsol &&  5.984 &  +0.1  &   3.074   & 71 \\
SCAN   &&  5.921 &  ~-0.8  & 3.341    & 12 \\
\hline 
\end{tabular}
\label{tab:lat}                                                                         
\end{table}

With the exception of LDA, the variations in lattice constants are relatively small across different XC functionals. However, the behavior of SCAN stands out distinctly from other GGA-based XC functionals on three key parameters. Firstly, it is commonly known that GGAs tend to overestimate lattice constants due to the underestimation of electron-electron correlations (Hartree and Hartree-Fock electron-electron correlations), while LDA exhibits the opposite trend. However, the meta-GGA SCAN demonstrates a unique behavior, leading to a decrease in lattice constant by approximately 0.8\%, contrasting with other GGA functionals. Secondly, a closer examination of the Fe magnetic moments in Table~\ref{tab:lat} reveals that $\mu^\text{Fe}_B$ is significantly enhanced (approximately 8\%) in the case of SCAN compared to PBE. This value is notably closer to the experimentally observed 3.3 $\mu_B$ from the only neutron diffraction study in Ref.~\cite{ferh_neutron}. Thirdly, the energy difference per formula unit with the FM state (E$_\text{FM}$-E$_\text{AFM}$) in Table~\ref{tab:lat}, calculated using optimized structures of respective XC functionals, is several times smaller for SCAN compared to other XC functionals. The AFM-FM transition temperature of 370 K ($\sim$ 32 meV) is slightly higher than the value E$_\text{FM}$-E$_\text{AFM}$ = 12 meV estimated by SCAN. For other functionals, this difference is higher than the transition temperature, suggesting that SCAN might be closer to describing the AFM-FM transition in FeRh. At this stage, we can assert that SCAN exhibits distinctive behavior for our current system of interest, FeRh. However, it remains unclear whether its performance is superior or inferior to other functionals and, in either case, how the results differ at a quantitative level. Addressing these questions requires further in-depth analysis, marking the next step in our ongoing study.

\begin{figure}[ht]
\centering
\includegraphics[width=7.5 cm]{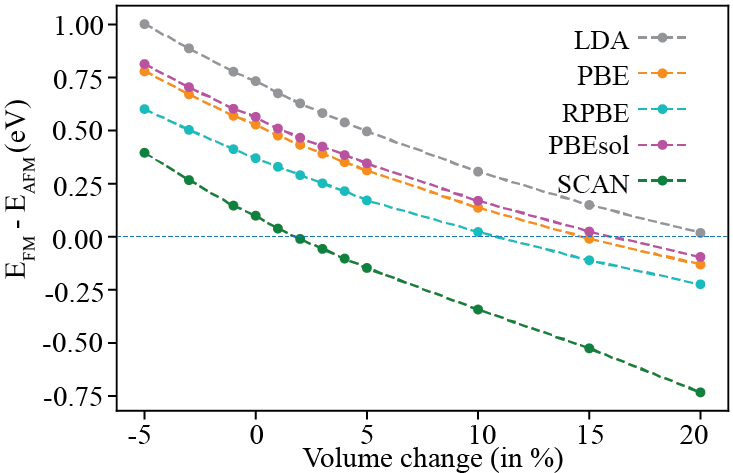}
\caption{Change in the magnetic ground state with the change of volume calculated for optimized crystal structure with various XC functionals. E$_\text{FM}$-E$_\text{AFM}$ infers AFM-FM transition. Notably, except SCAN, all other functionals anticipate the transition occurring at a substantially higher volume than the experimentally observed approximately 1 \% change.}
  \label{fig:fig01}
\end{figure} 

As the next step in our analysis, we calculate the energy difference between the FM and AFM states (E$_\text{FM}$-E$_\text{AFM}$) with the change in volume using different XC-functionals. The corresponding plot is presented in Fig.~\ref{fig:fig01}. Notably, the AFM-FM transition occurs at approximately 2 \% volume expansion for SCAN, while for PBE/PBEsol, it takes place after a $\sim$ 15 \% expansion. RPBE exhibits relatively better performance among GGAs, although the transition still occurs after a substantial $\sim$ 10 \% volume expansion. Hence, at first glance, SCAN appears to better perform compared to other functionals, particularly in describing the magnetostructural transition in the FeRh system. Consistent with previous DFT-based studies, in the FM state, Rh magnetic moments at the pristine volume are $\sim$ 1.061 $\mu_B$ while no net Rh moment is observed in the AFM state.  The appearance/disappearance of Rh moments is believed to be attributed to hybridization among Fe and Rh atoms ~\cite{dispute8, dispute9}. The Fe moment remains largely unchanged throughout the transition.

\begin{figure}[ht]
\centering
\includegraphics[width=7.5 cm]{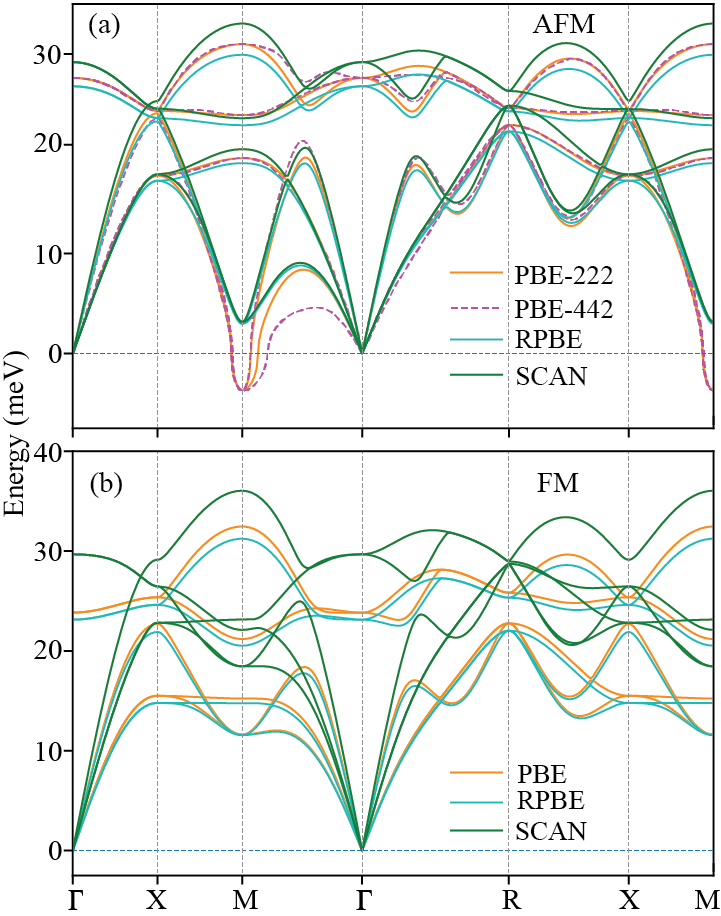}
\caption{Phonon dispersion is computed for both the (a) AFM and (b) FM phases utilizing the finite displacement method with PBE, RPBE, and SCAN functionals. For the PBE case in (a), we employ two unit cells of sizes 2 $\times$ 2 $\times$ 2 and 4 $\times$ 4 $\times$ 2 to ensure convergence concerning unit cell size.}
  \label{fig:fig02}
\end{figure} 

To further assess the reliability of the SCAN functional, we evaluate the stability of crystal structures by computing phonon dispersion. A similar analysis is conducted for comparison purposes using PBE and RPBE functionals. The resulting plots for the AFM and FM phases are presented in Fig.~\ref{fig:fig02}(a) and Fig.~\ref{fig:fig02}(b) respectively. Additional details of the calculations are available in the Methodology section.  The phonon dispersion of the two phases $viz$ FM and AFM, exhibits notable differences, particularly with the softening of optical modes in the AFM phase. This observation underscores the pronounced influence of spin-lattice coupling in the FeRh system. In the FM phase, all three functionals-PBE, RPBE, and SCAN—consistently predict the dynamical stability of the crystal lattice. Interestingly, soft phonon modes are observed with the PBE functional at the M point for the AFM phase in Fig.~\ref{fig:fig02}(a), a feature absent in the case of RPBE and SCAN functionals. Similar soft modes have been reported in prior DFT-based alculations~\cite{dft_lat5, dft_lat6}. In support of this, Wolloch {\it et. al.}~\cite{dft_lat6} argued that large Fe magnetic moments can dynamically stabilize the crystal structure in the AFM phase. Both Aschauer {\it et. al.}~\cite{dft_lat5} and Wolloch {\it et. al.}~\cite{dft_lat6} manipulated Fe magnetic moments to reach this conclusion, either through the DFT + U approach or by artificially increasing them. In our current study, both RPBE and SCAN functionals (RPBE functional in Ref.~\cite{dft_lat6}) demonstrate success in predicting the dynamically stable AFM phase of FeRh. However, only SCAN appears to reasonably predict the AFM-FM transition with volume change. This discrepancy prompts a natural question about the differences at microscopic level from these two functionals in describing the properties of the FeRh alloy.

\begin{figure*}[ht]
\centering
\includegraphics[width=16 cm,height = 4 cm]{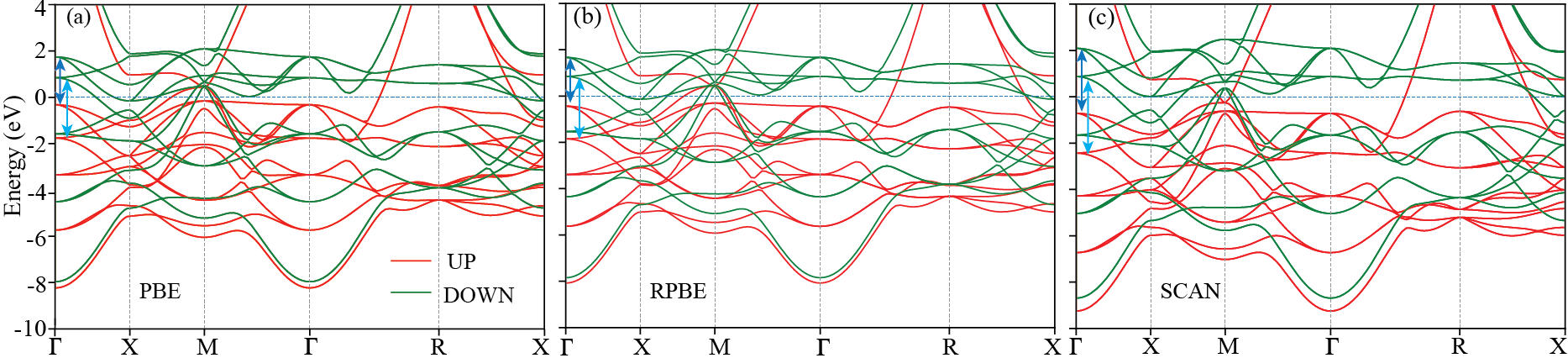}
\caption{Spin-polarized band structure plots for the FM phase are presented with (a) PBE, (b) RPBE, and (c) SCAN functionals. We used two-atom primitive unit cell in our calculations. Bands in the up and down spin channels are depicted in red and green, respectively, with the Fermi level set to zero in the plots. The two differently colored arrows around Fermi level in the plots illustrate the exchange splitting of the top two bands, measuring 2.064 and 2.603 eV from PBE in (a), 2.105 and 2.699 eV from RPBE in (b), and 2.809 and 3.307 eV from SCAN in (c).}
  \label{fig:fig03}
\end{figure*}

To address the aforementioned point, we conduct a spin-polarized band structure analysis using PBE, RPBE, and SCAN functionals for the FM phase. In this investigation, we consider the primitive cell consisting of two atoms to provide a clear observation of the impact of various functionals on the electronic structure. The corresponding plots are presented in Fig.~\ref{fig:fig03}. Two significant observations arise from this analysis, shedding light on the distinctive behavior of SCAN compared to the two GGA functionals. Firstly, owing to its improved handling of delocalization errors, SCAN pushes the occupied bands downward and the unoccupied bands upward relative to the Fermi level (Fig.~\ref{fig:fig03}(c)), contrasting with the band structures obtained with PBE and RPBE (which exhibit more or less the same behavior) as shown in Fig.~\ref{fig:fig03}(a) and (b). This behavior of SCAN suggests a commendable treatment of electronic correlation, in particular of the Fe/Rh $d$ orbitals contributing dominantly (not explicitly shown) to the band structure plots in Fig.~\ref{fig:fig03}. Secondly, and intricately connected with the first observation, SCAN reveals an augmented exchange splitting at the Fermi level. This enhancement is highlighted by arrows with two different colors for the top two bands in the plots. In the case of PBE and RPBE functionals, the splitting—depicted by dark and light blue arrows in Fig.~\ref{fig:fig03}(a),(b)-measures 2.064, 2.603 eV and 2.105, 2.699 eV, respectively. In contrast, the splitting increases to 2.809 and 3.307 eV with the SCAN functional in Fig.~\ref{fig:fig03}(c). The disparity in exchange splitting between these GGA functionals and SCAN aligns closely with the error by which PBE/RPBE exaggerates the AFM phase stability over the FM phase in Fig.~\ref{fig:fig01}. Consequently, one can attribute the relatively superior description of the AFM-FM transition in FeRh by the SCAN functional to its capacity to bring a substantial exchange splitting in the electronic structure.

The seemingly promising findings regarding the performance of the SCAN functional for the FeRh system, particularly concerning phonon dispersion and the AFM-FM transition with volume, require further examination. To validate the efficacy of the SCAN functional in accurately describing FeRh, magnetic interactions obtained with it should be able to reproduce the experimentally observed paramagnetic transition at $\sim$ 670 K. We undertake a thorough investigation of this aspect in following two steps.

\begin{figure}[ht]
\centering
\includegraphics[width=8 cm]{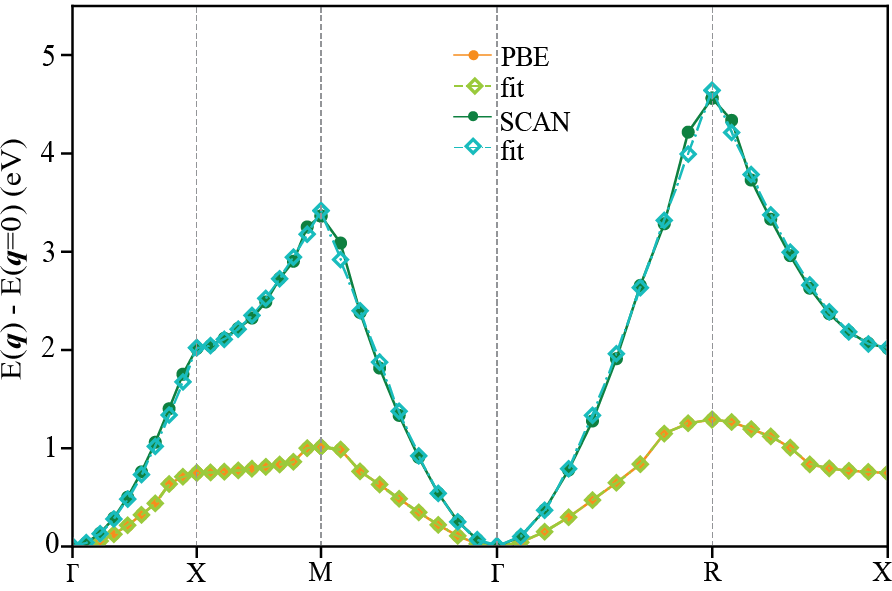}
\caption{Spin-spiral calculations are performed for the low-temperature AFM phase using PBE (orange-filled circles) and SCAN (green-filled circles) XC functionals. In this representation, the $\Gamma$ point signifies the AFM ground state. The obtained data is subjected to fitting with the Heisenberg model given in Eq. (1), involving up to 4$^{th}$ Fe-Fe neighbors, using linear regression. The results of the fitting are depicted in the plot with light-green and blue empty diamond symbols for PBE and SCAN functionals, respectively.}
  \label{fig:fig04}
\end{figure}

In the first step, the spin-spiral calculations as the starting point of Frozen Magnon Approximation are performed. As explained in the Methodology section, each spin spiral state characterized by $\bm{q}$ is equivalent to a magnetic configuration with energy $E(\bm{q})$. Fig.~\ref{fig:fig04} depicts the stabilization energy ($E(\bm{q})$ - $E(\bm{q}$ = 0) dispersions for various noncollinear magnetic configurations along the $\Gamma -X-M- \Gamma -R-X$ path, calculated using both PBE and SCAN XC-functionals. In this plot, $\Gamma$-point ($\bm{q}$ = 0) represents the AFM state observed at low temperatures. To model the system within a Heisenberg model, linear regression was employed to extract Fe-Fe magnetic exchange interactions up to the fourth nearest neighbor. Coefficient of determination ($R^2$) are found to be 0.99 and 0.98 for the PBE and SCAN XC-functionals respectively, indicating the model's ability to capture the energy dispersion accurately. This can further be qualitatively verified by looking at the goodness of fitting.

 Upon examining Fig.~\ref{fig:fig04}, two straightforward observations emerge. Firstly, from both functionals, the FeRh system attains its lowest energy at the $\Gamma$ point, indicating that the AFM state is the ground state. This again is consistent with experimental observation and highlight the fact that both these functional correctly obtain the magnetic ground state. Secondly, while the nature of the plots from PBE and SCAN functionals remains largely similar, there is a significant discrepancy in the energy scale. The relative energies [$E(\bm{q})$ - $E(\bm{q} = 0)$] are notably larger with SCAN compared to PBE. One way to look at this point is that SCAN predicts the AFM state at low tenperature to be robustly stabilized against any other competing magnetic orders, particular against the FM state represented by $R$ point in Fig.~\ref{fig:fig04}. However, from Table~\ref{tab:lat}, one can see that E$_\text{FM}$-E$_\text{AFM}$ = 12 meV meaning FM state is lying in close vicinity to that of the AFM state. This observation immediately prompts the question of whether their is also discrepancies among magnetic interactions obtained from the two functionals and, if so, which one of them aligns with the experimentally observed paramagnetic transition temperature of $\sim$ 670 K. To address these questions and assess the validity of SCAN and PBE for this challenging system, extraction of magnetic interactions and subsequent calculations of the magnetic ordering temperature are performed in the second step.

\begin{figure}[ht]
\centering
\includegraphics[width=6 cm]{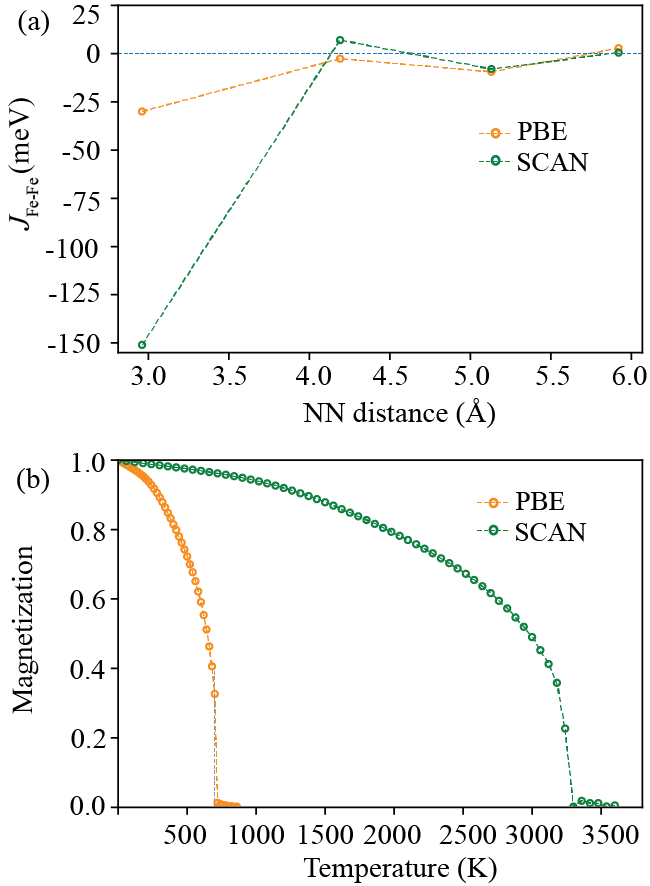}
\caption{(a) Magnetic interactions between Fe atoms in the AFM ground state are estimated using the Frozen-magnon approach, as explained in the methodology section, with both PBE and SCAN functionals. The 1$^{st}$ neighbor interaction is antiferromagnetic, consistent with the magnetic ground state. (b) The variation of normalized magnetization with temperature is illustrated in our Monte Carlo simulations. The interactions obtained with PBE in (a) yield a magnetic ordering temperature $T_\text{N}$ $\sim$ 700 K, closely matching an experimental estimation of 670 K. Conversely, the magnetic interactions from the SCAN functional significantly overestimate it.}
  \label{fig:fig05}
\end{figure}

Hence, to elucidate the impact of XC functionals, we first extract Fe-Fe magnetic interactions ($J$'s) up to the fourth nearest neighbor using the inverse Fourier transform method within the Frozen Magnon approach, discussed in the Methodology section. Subsequently, Monte Carlo simulations are employed to estimate the magnetic ordering temperature. Fig.~\ref{fig:fig05}(a) depicts the extracted interactions plotted against different nearest neighbor distances. Both PBE and SCAN agree on the dominant antiferromagnetic first-neighbor interaction ($J_1$) at $\sim$ 2.96 $\AA$, consistent with the observed antiferromagnetic ground state.  However, their magnitudes differ significantly. SCAN predicts a substantially larger $J_1$ of -151 meV compared to -30 meV predicted by PBE. Interestingly, while smaller, the second-neighbor interaction ($J_2$) at 4.19 $\AA$ exhibits a contrasting behavior. SCAN predicts a ferromagnetic $J_2$ of 7 meV, whereas PBE estimates it as antiferromagnetic with a value of -2.6 meV. 
 In the low-temperature AFM configuration, first and second-neighbor couplings are expected to be AFM and FM, respectively. Therefore, SCAN's FM $J_2$ at first glance appears more consistent. The estimated third-neighbor interactions $J_3$ at 5.13 $\AA$ are -8 and -9.5 meV, and the fourth-neighbor interactions $J_4$ at 5.92 $\AA$ are 0 and 2.8 meV for SCAN and PBE, respectively. These discrepancies highlight the significant qualitative differences in magnetic interactions extracted using the two XC functionals.

To check upon the reliability of the two set of interactions obtained above in reproducing experimentally observed magnetic ordering temperature, we perform Monte Carlo simulations considering the Heisenberg model on the simple cubic lattice of Fe atoms (see methodology). Obtained variation in normalized as a function of temperature is plotted in Fig.~\ref{fig:fig05}(b) for the two cases, $viz$ PBE and SCAN. The estimated magnetic ordering temperature with $J$'s estimated using  PBE and SCAN functionals $\sim$ 710 K and 3240 K respectively. Though this temperature is overestimated in both cases, the case of SCAN is in complete disagreement with the experimentally observed value of $\sim$ 670 K. It an order of magnitude overestimated when compared to PBE-obtained value which is in a better agreement with experiment. This discrepancy can directly be attributed to large $J_1$ obtained from SCAN in Fig.~\ref{fig:fig05}(a). As far the PBE is concerned, one can tune $J$'s through the application of $U$ on Fe-$d$ states within the PBE+$U$ approach to achieve a closer agreement with experimentally observed magnetic ordering temperature. However, our presented results suffice to highlight the behavior of different XC functional in describing structural, electronic, and magnetic properties of the FeRh system.   


\section{Conclusion}
In conclusion, we have investigated the performance of various XC functionals within the DFT formalism for the challenging binary alloy FeRh. Employing LSDA, various GGAs, and newly developed SCAN meta-GGA functional, our results show the limitations of any single functional in capturing the intricate interplay of structural, electronic, and magnetic properties in FeRh. We have shown that while SCAN can accurately describe some magnetic features and phonon dispersion, it significantly overestimates the Fe-Fe magnetic interactions, leading to an unreasonable magnetic ordering temperature. Conversely, the PBE GGA exhibits the opposite behavior than SCAN. These findings highlight the challenges in simulating materials with partially filled $d$-shells using DFT, underscoring the crucial need for developing a versatile XC functional that can effectively account for the multifaceted nature of such systems.


\bibliographystyle{elsarticle-harv}

\end{document}